\documentclass[aps,prd,amsmath,amsfonts,amssymb,eqsecnum,nofootinbib,
  floatfix,secnumarabic,preprintnumbers,superscriptaddress,nobalancelastpage,
  onecolumn,notitlepage,biblatex]{revtex4-1}
\usepackage[utf8]{inputenc}
\usepackage{color,graphicx,hyperref,dsfont,slashed,multirow,subfig}

\hypersetup{bookmarks=true,unicode=true,pdftoolbar=true,pdfmenubar=true,
  pdffitwindow=false,pdfstartview={FitH},pdftitle={HeavyHiggs},
  pdfauthor={B.~Coleppa,~B.~Fuks,~P.Poulose,~S.~Sahoo},
  pdfsubject={Subject},pdfcreator={Creator},pdfproducer={Producer},
  pdfkeywords={Heavy Higgs bosons}{2HDM},pdfnewwindow=true,colorlinks=true,
  linkcolor=blue,citecolor=magenta,filecolor=magenta,urlcolor=cyan}
\newcommand{\met}{\ensuremath{{\not\mathrel{E}}_T}}

\def\bsp#1\esp{\begin{split}#1\end{split}}
\def\be{\begin{equation}}
\def\ee{\end{equation}}
\def\bpm{\begin{pmatrix}}
\def\epm{\end{pmatrix}}

\begin{document}

\title{Fermiophobic gauge boson phenomenology in 221 Models}

\author{Baradhwaj Coleppa}
\email{baradhwaj@iitgn.ac.in}

\author{Satendra Kumar}
\email{satendrak@iitgn.ac.in}

\author{Agnivo Sarkar}
\email{agnivo.sarkar@iitgn.ac.in}
\affiliation{Indian Institute of Technology Gandhinagar,Gandhinagar, India}


\begin{abstract}
Models with extra gauge symmetry are well-motivated extensions of the Standard Model. In this paper, we study an extended gauge model with a heavy neutral gauge boson $Z'$ which is fermiophobic. Thus, the production of such particles can occur via vector boson fusion, with subsequent decays into $WW$ or $Zh$. We investigate the collider phenomenology of such $Z'$s in the context of both the 14 TeV LHC and the future CLIC. We find that looking at $\ell\ell b b$ final states provides a rich opportunity to discover such new vector bosons where conventional search strategies in the dilepton channel would fail. In particular, we optimize our analysis by putting in kinematic cuts deriving model-independent values of $\sigma\times$BR needed for a 5$\sigma$ discovery at the LHC. We then translate this into the parameter space of a specific model for illustration purpose -- our results show that fermiophobic $Z'$s are discoverable in the $\ell\ell b b$ channel for wide range of parameter values.
\end{abstract}

\maketitle

\section{Introduction}\label{sec:intro}
One of the primary goals in particle physics is to understand the precise mechanism responsible for Electroweak Symmetry Breaking (EWSB). The Standard Model (SM) engineers EWSB via the introduction of a Higgs doublet that develops a vacuum expectation value (vev) that spontaneously breaks the gauge symmetry from $SU(2)\times U(1)$ down to electromagnetism. The evidence in favor of the SM, already enormous, has received a boost with the discovery of a scalar particle that has properties consistent with those of the SM Higgs  \cite{Aad:2012tfa, ATLAS:2013sla, Chatrchyan:2012ufa,CMS:yva, Chatrchyan:2014vua, Chatrchyan:2013zna}, and its mass and spin are now known \cite{Aad:2013xqa}. At the same time, there remain vexing questions about the large hierarchy between the Planck and the electroweak scales, a more ``natural" explanation for the Yukawa mass terms of the fermions, origin of dark matter etc. - questions that are not answered within the SM. Over the years, many models purporting to go ``beyond the SM" (BSM) have appeared in the literature - a common feature of these models is the presence of some new physics around the TeV scale either in the form of new heavy vector, scalar, or fermion resonances. Understanding the phenomenology of these models at the LHC is currently a priority.

One of the earliest BSM scenarios to emerge was Technicolor (TC) - a new strong force with dynamics similar to QCD that breaks the electroweak symmetry dynamically generating $W$ and $Z$ boson masses. Along with its many avatars (Extended TC, Top Color Assisted TC), these models have been studied well and their features documented \cite{Hill:1991at,Hill:1994hp,Hill:2002ap}. With the discovery of the AdS-CFT correspondence \cite{Maldacena:1997re,Gubser:1998bc}, it later emerged that this class of strongly coupled models have a weakly coupled extra dimensional dual description. The advantage of using the AdS-CFT prescription was that these extra dimensional models now permitted a perturbative study. Being dual to TC, these models did not rely on the existence of a fundamental scalar particle in the spectrum and hence were dubbed ``Higgsless models" \cite{Csaki:2003dt} - these were realized on a slice of $AdS_{5}$ with the symmetry breaking being encoded by the boundary conditions. Using the idea of ``dimensional deconstruction"\cite{ArkaniHamed:2001ca}, these extra dimensional gauge theories could be understood as a collection of 4D gauge groups connected together by non-linear sigma model fields. This picture is called a ``Moose" or ``Quiver" diagram \cite{Georgi:1985hf}. The simplest realization of such models that relied on just one extra gauge group was presented in \cite{Chivukula:2006cg,Chivukula:2009ck} - in addition to the SM spectrum, this model contained an extra set of vector resonances $W',Z'$ and vector fermion partners for every species of SM quarks and leptons. 

The discovery of a Higgs boson necessarily implies that purely Higgsless models cannot be the whole story. To this end, Ref.~\cite{Abe:2012fb} constructed a UV completion of the three site model introducing two Higgs doublets to break the gauge symmetries. This model shares many of the features pertaining to EWSB with other $SU(2)\times SU(2)\times U(1)$ models (which we will call 221 Model henceforth) - in addition it has certain distinguishing features in that the $W',Z'$ in this model are fermiophobic resulting in markedly new ways of production and detection compared to other 221 models. In the context of the original Higgsless model, this has been studied in Ref.~\cite{Du:2012vh} - the goal of the present study is to extend this work within the context of the ``Higgsful" 221 model of Ref.~\cite{Abe:2012fb}. Other studies of $Z'$ phenomenology (not necessarily in fermiophobic models) can be found in the literature for example in \cite{Rizzo:2006nw,Langacker:2008yv,Bandyopadhyay:2018cwu}.

This paper is organized as follows: In Sec.~\ref{sec:221 model}, we present the necessary details of the model under investigation. Sec.~\ref{sec:pheno} starts with the production mechanism and classification of the decays of the $Z'$ in this model. We also briefly present the current bounds, and in Sec.~\ref{subsec:LHC}, we present the LHC analysis picking certain benchmark points. We move on to understand the discovery potential of the $Z'$ at the future CLIC collider in Sec.~\ref{subsec:CLIC}, and in Sec.~\ref{sec:conclusions}, we present our conclusions.


\section{The 221 model: Higgsless to Higgsful}
\label{sec:221 model}
In this section, we present an overview of the specific model (hereafter referred to as the 221 model) that we would use as an operating example for our phenomenological model later. The discussion here is very brief, only touching upon the parts directly relevant to us -- for a complete description of the model the reader may consult Refs.~\cite{Chivukula:2006cg},\cite{Abe:2012fb}.

\begin{figure} [h!]
\vspace{0.2in}
\includegraphics[angle=0,width=80mm]{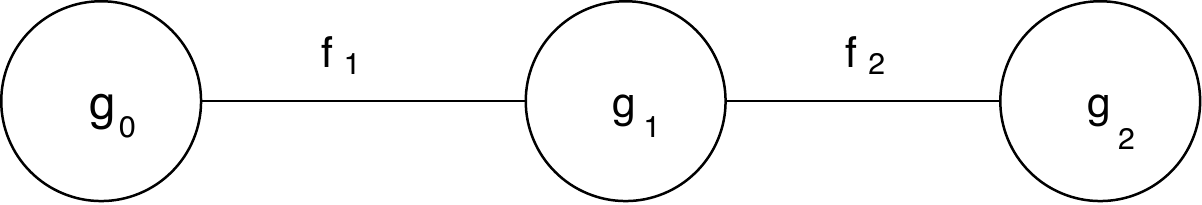}
\vspace{-2mm}\caption{The first two circles represent $SU(2)$ gauge groups of coupling strengths $g_0$ and $g_1$, while the third is a $U(1)$ gauge group with coupling strength $g_2$.  There are two Higgs doublets $\Phi_1$ and $\Phi_2$ connecting sites 0-1 and 1-2 respectively whose vacuum expectation values are denoted by $f_1$ and $f_2$.}
\label{fig:3smm}
\end{figure}

The gauge sector of the 221 model is $SU(2)\times SU(2)\times U(1)$ -- this is completely broken down to $U(1)_{\textrm{em}}$ when two Higgs fields $\Phi_1$ and $\Phi_2$ develop vev's $f_1$ and $f_2$. The model is conveniently described by a "moose" diagram as depicted in Fig.~\ref{fig:3smm}. The first and second "sites" are the two $SU(2)$ groups characterized by their couplings $g_0$ and $g_1$, while the third is the $U(1)$ group with coupling constant $g_2$. The left-handed fermions are represented by the lower vertical lines and the right-handed ones by the upper ones. The model admits fermion mass terms via the usual Yukawa coupling of the form $\bar{\Psi}_{L0}\Phi_1\Psi_{R1}$ etc and also a Dirac mass term of the form 
$\bar{\Psi}_{L1}\Psi_{R1}$. We will not discuss the mechanism of fermion mass generation in the model further in this paper - the interested reader is referred to \cite{Abe:2012fb}.

The gauge and scalar sectors of are described by the Lagrangian:

\begin{eqnarray}
{\cal L}&=&-\frac{1}{4} \sum\limits_{a=1}^3 W^a_{0\mu\nu}W_a^{0\mu\nu} -\frac{1}{4} \sum\limits_{a=1}^3 W^a_{1\mu\nu}W_a^{1\mu\nu} -\frac{1}{4} B_{2\mu\nu}B^{2\mu\nu} \nonumber\\
&&+ \sum\limits_{j=1,2} \textrm{Tr}\Big[(D_\mu \Phi_j)^{\dagger}(D^\mu \Phi_j)\Big] - V(\Phi_1, \Phi_2)
\label{eqn:LagGH}
\end{eqnarray}
where $W_0^{a\mu\nu}, W_1^{a\mu\nu}$ and $B_2^{\mu\nu}$ are the field strengths corresponding $SU(2)_0, SU(2)_1$ and $U(1)_2$ gauge groups. The most general form of Higgs potential $V(\Phi_1, \Phi_2)$ that respects the gauge symmetry is given by

\begin{eqnarray}
V(\Phi_1, \Phi_2) &=& \frac{1}{2} \lambda_1 \Big[\textrm{Tr}(\Phi_1^\dagger \Phi_1) - \frac{f_1^2}{2} \Big]^2+\frac{1}{2} \lambda_2 \Big[\textrm{Tr}(\Phi_2^\dagger \Phi_2) - \frac{f_2^2}{2} \Big]^2 \nonumber\\
&&+ \lambda_{12} \Big[\textrm{Tr}(\Phi_1^\dagger \Phi_1) - \frac{f_1^2}{2} \Big] \Big[\textrm{Tr}(\Phi_2^\dagger \Phi_2) - \frac{f_2^2}{2} \Big].
\label{eqn:LagH}
\end{eqnarray}
Writing the two Higgs field in the matrix form:
\begin{eqnarray}
\Phi_j &=& \frac{1}{2}\big( f_j + h_j + i \tau^a \pi_j^a \big), ~~~~~~~~~~ j = 1,~2
\end{eqnarray}
the mass matrix for the CP-even state ($h_1, h_2$) can be diagonalized to yield the states:

\begin{eqnarray}
h &=& \cos\alpha\, h_1-\sin\alpha\, h_2 \nonumber\\
H &=& \sin\alpha\, h_1+\cos\alpha\, h_2.
\label{eqn:higgs_states}
\end{eqnarray}

The kinetic energy terms of the $\Phi_j$'s are governed by the covariant derivatives given by:
\begin{eqnarray}
D_\mu \Phi_1 &=&  \partial_\mu\Phi_1+ig_0 \frac{\tau^a}{2} W_{0\mu}^a \Phi_1-ig_1\Phi_1 \frac{\tau^a}{2} W_{1\mu}^a  \\
D_\mu \Phi_2 &=&  \partial_\mu\Phi_2+ig_1 \frac{\tau^a}{2} W_{0\mu}^a \Phi_2-ig_2\Phi_2 \frac{\tau^3}{2} B_{2\mu}
\label{eqn:cova}
\end{eqnarray}
When the two Higgs fields develop vevs, the kinetic terms yield, among other things, mass terms for the gauge bosons - the structure of the mass matrix guarantees one zero eigenvalue which is identified as the photon. The eight degrees of freedom in the scalar sector combine to make the $W^{\pm},Z,W'^{\pm}$ and $Z'$ massive - the remaining two degrees of freedom are identified with the physical particles given in Eqn.~\ref{eqn:higgs_states}.

Of particular interest to us is that the $Z'$, being fermiophobic \cite{Foadi:2004ps} \cite{Chivukula:2005xm}, only decays to $Zh$, $ZH$, and $WW$ -- below we give the partial decay widths to $Zh$ and $WW$:
\begin{eqnarray}
\Gamma (Z' \rightarrow Z h)=&& \frac{g_{Z'Zh}^{2} v^2 M_{Z'}}{786 \pi M_Z^2}\Bigg[2+\frac{(M_{Z'}^2+m_Z^2-m_h^2)^2}{4m_Z^2 M_{Z'}^2}\Bigg]\times\sqrt{\Big[1+ \frac{(m_h^2-m_Z^2)^2}{M_{Z'^4}}-2\frac{(m_h^2+m_Z^2)}{M_{Z'^2}}}\Big] \\
\nonumber\\
\Gamma (Z' \rightarrow W W)=&& \frac{g_{Z'WW}^{2} M_{Z'}^5}{192 \pi M_W^4}\sqrt{\Big[1-\frac{4 m_W^2}{M^2_{Z'}}\Big]}\times \Bigg[1+16\frac{M_W^2}{M_{Z'}^2}-68\frac{M_W^4}{M_{Z'}^4}-48\frac{M_W^6}{M_{Z'}^6}\Bigg]. \\\nonumber 
\label{eqn:Decay}
\end{eqnarray}
The couplings $g_{Z'WW}$ and $g_{Z'Zh}$ are determined in terms of the parameters $r=f_2/f_1$ and $x=\frac{1+r^2}{r}\frac{m_W}{m_{W'}}$. In the phenomenological analysis to follow, we obtain discovery regions of the $Z'$ in the plane of $g_{Z'WW}$-$g_{Z'Zh}$ setting $r=1$.


\section{Collider Phenomenology}
\label{sec:pheno}

\subsection{Cross-sections, Decay Rates and $Z'$ search strategy}

We begin our feasibility studies of a fermiophobic $Z'$ discovery at the LHC and CLIC by first setting up the calculation and the specific search strategy employed. While the assumption that the $Z'$ is fermiophobic automatically does imply a model-dependent analysis, we will present the analysis in a more general framework without recourse to a particular model in Sections \ref{subsec:LHC} and \ref{subsec:CLIC}, and then interpret these results within the context of the model described in Section \ref{sec:221 model}. The first step in this analysis is thus the identification of the production and decay modes of the $Z'$ -- typically, in the absence of fermionic couplings, the production of $Z'$ proceeds mostly via vector boson fusion (VBF). While there could be models in which this might not necessarily be the dominant production mechanism, we restrict our analysis to this production channel as it is sufficiently general. The decay, however, can proceed via multiple mechanisms depending on the strength of couplings in a particular model, and the availability of phase space for the benchmark points chosen. In what follows, we consider the decay $Z'\to Zh$, where the $h$ could be the SM-like higgs, or any other (heavier) scalar particle in the theory. Thus the complete channel chosen to study the prospects of discovery is $pp\to Z'\to Zh \to bb\ell\ell$, with the higgs decaying to a pair of bottom quarks and the $Z$ decaying leptonically. While the hadronic decays of $Z$ could certainly be considered, we find that this channel offers maximum reach in spite of the suppressed branching ratios as the presence of the two leptons in the final state greatly reduces the pure QCD background.

\begin{figure}[!ht]
\begin{center}
\includegraphics[angle=0,width=80mm]{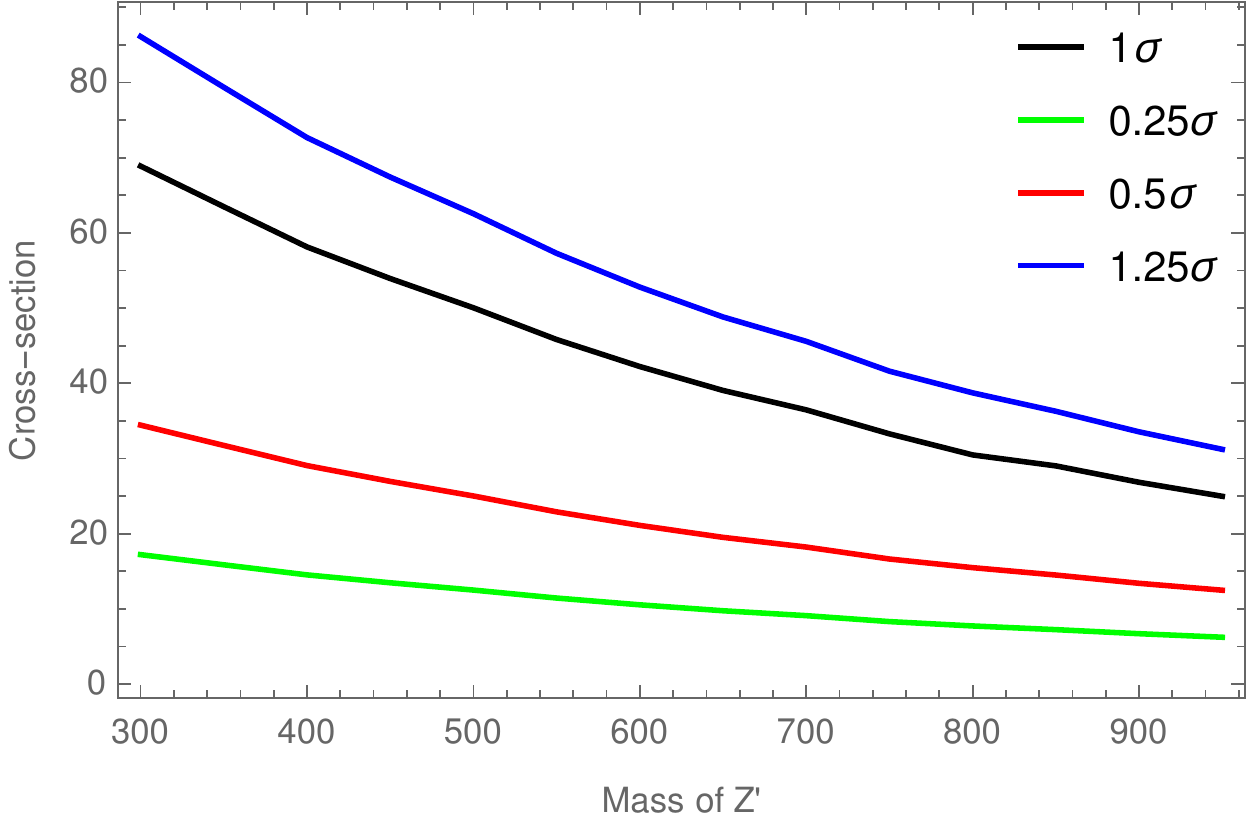}
\caption{The cross section across $M_{Z'}$ for the $Z'jj$ production at fixed center-of-mass energy of 14 TeV assuming the corresponding SM value for the coupling $g_{Z'WW}$ (black curve). We have also shown how the cross-section would change for $g_{Z'WW}^2/g_{\rm{SM}}^2=$0.5, 0.25, and 1.25.}
\label{fig:CS}
\end{center}
\end{figure}

\begin{figure}[!ht]
\begin{center}
\includegraphics[angle=0,width=80mm]{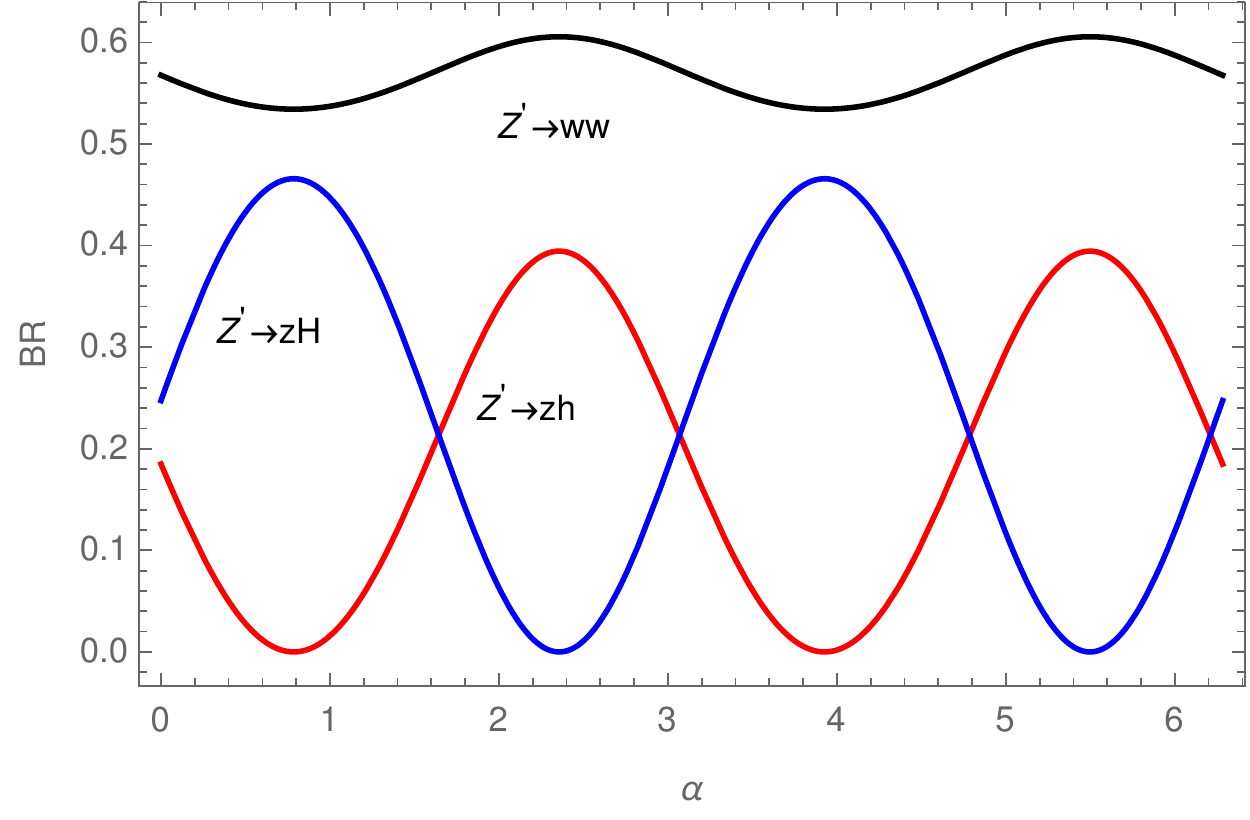}
\caption{The branching ratios of the the $Z'$ to the final states $Zh$, $ZH$, and $WW$ as a function of the mixing angle in the scalar sector in the context of the  221 model considered in the previous section. While the $WW$ decay generically dominates, it is seen that the decay into $Zh$ final state is non-negligible for a wide range of parameter values.}
\label{fig:BR}
\end{center}
\end{figure}

While the production cross-section depends on the coupling $g_{Z'WW}$, to get an idea of the numbers involved, we reproduce the plot in  \cite{Bolzoni:2011cu} that calculates the cross-section at the 14 TeV LHC assuming the corresponding SM value for the coupling in Fig.~\ref{fig:CS}. 
We have also shown in the plot curves corresponding to $g_{Z'WW}^2/g_{\textrm{SM}}^2=$ 0.25, 0.5, and 1.25 using simple scaling. It is seen that the cross-sections for the production of  a moderately heavy $Z'$ in the range 300 GeV - 1 TeV is in tens of pb -- while this is certainly good news, one need also consider the branching ratios involved. While we will choose $M_{Z'}=$300, 500, and 700 GeV as our benchmark points in the analyses to follow, we present in Fig.~\ref{fig:BR} the plots of the branching ratios involved in the 221 model for $M_{Z'}=$500 GeV for illustration purposes -- it is clearly seen that while $WW$ can dominate, the $Zh$ and $ZH$ are non-negligible over a wide range of parameter spaces and are thus important channels to consider.

In what follows, we will present the analysis for the LHC and CLIC --  while we will choose a particular model for this part, we stress here that this is only to analyze the efficacy of the set of cuts we will devise. The signal cross-section we will choose is arbitrary. We will look at the complete set of SM background and devise an optimal set of cuts that will suppress the SM without affecting the signal too much. We will then back-calculate the signal cross-section necessary for a 5$\sigma$ discovery and translate this number into the parameter space of the 221 model to understand the reach.

\subsection{Experimental limits}

The search for a heavy neutral heavy gauge boson has been performed at both the Tevatron and the LHC. The predominant decay channel in most of these searches is the $\ell^{+}\ell^{-}$ final state and thus these limits do not apply in our case as we are considering fermiophobic $Z'$s. However, we present below a pr\'ecis of the various searches performed for the sake of completeness. The Tevatron looked for the  $Z'$ in the channel ($p\bar{p} \rightarrow Z^{'} X \rightarrow l^{+} l^{-} X$) \cite{Carena:2004xs}, where it is assumed that the $Z^{'}$ is produced via light quark coupling. The analysis performed after collecting the data did not show any significant excess. The experiment done by CDF and D0 collaboration translated their result for various $U(1)$ extensions of the SM. The result excluded $Z'$s in mass range 500-800 GeV, depending upon model specifications.

At the LHC, the CMS collaboration did extensive searches for the $Z^{'}$ both at 8-TeV\cite{Khachatryan:2014fba} and 13-TeV\cite{Khachatryan:2016zqb} c.m.s energy scale. For the 8-TeV case $Z'$ has been searched for integrated luminosity 20.6$fb^{-1}$(19.7$fb^{-1}$) in the final states $e^{+}e^{-}$($\mu^{+}\mu^{-}$) and the negative result obtained was used to set a 95\% C.L. upper limit on production cross-section $\times$branching ratio of $Z^{'}$ to $\ell\ell$. This result when translated to various BSM models excludes $Z'$  up to a TeV scale. The search performed at CMS-13\cite{Khachatryan:2016zqb} followed a similar strategy and the limits are understandably stronger. Apart from the resonant decay through dilepton channel, ATLAS collaboration \cite{Aaboud:2016cth},\cite{Aaboud:2017buh} also searched for non-resonant decay channel. We summarize the results of the experimental searches in Table ~\ref{tab:limits}. As mentioned previously, in the analysis to follow we treat $M_{Z'}$ as essentially a free parameter choosing 300, 500, and 700 GeV as our benchmark points as none of the search channels with dilepton final states apply in our case.

\begin{table}[h!]
\begin{tabular}{|c|c|c|c|c|c|c|c|}
\hline
 Collaboration   &Luminosity($fb^{-1}$)  &$Z^{'}_{SSM}$  &$Z^{'}_{\psi}$ &K-K Mode &K-K Mode &$Z^{'}_{\kappa}$   &$l\bar{l}q\bar{q}$ (Non-resonant decay) \\
& & &  &0.01  &0.1 &  & \\ 
\hline\hline
 CMS-8                                        &20.6(19.7)  &2.9 TeV  &2.577 TeV &1.27 TeV &2.73 TeV & - & -  \\
CMS-13                                       &2.7(2.9)   &3.37 TeV   &2.82 TeV &1.46 TeV  &3.11 TeV  &- &- \\
ATLAS-13                                       &3.2   &3.36 TeV    &2.74 TeV   &-      &-  &2.74 TeV &16.7 TeV - 25.2 TeV \\
ATLAS-13                           &36.1   &4.5 TeV    &3.8 TeV & -      & - & 4.1 TeV & 24 TeV - 40 TeV \\
\hline
\end{tabular}
\caption{Excluded regions in $M_{Z'}$ for various models on the basis of the searches at the LHC thus far. It is seen that for all models that allow $Z'$ couplings to fermions, the experiments already set very stringent lower limits for $M_{Z'}$.}
\label{tab:limits}
\end{table}

\subsection{$Z^{'}$ search prospects at the LHC}
\label{subsec:LHC}

We now present the analysis for the search of the $Z'$ in the $\ell\ell bb$ channel at the 14 TeV LHC. To do so, we start by choosing as our benchmark points $M_{Z'}=$300, 500, and 700 GeV.  The $Z'$ is presumed to decay to $Zh$, with the $h$ being the SM-like higgs. There is no \emph{a priori} reason why this higgs need be the SM one and not a heavier state itself, but in order to make the analysis simpler, we have chosen it to be the already discovered boson with well measured BR values to $b\bar{b}$.  The data simulation is performed using {\sc MadGraph}5\_{aMC@NLO} \cite{Alwall:2014hca} event generator with the center of mass energy fixed at 14 TeV. While the SM background events are generated using the inbuilt SM model file in the {\sc MadGraph} repository, the 221 model was built using the {\sc FeynRules} program \cite{Degrande:2011ua, Christensen:2009jx}. Parton level events generated are from {\sc MadGraph} are then passed on to {\sc Pythia}~6~ \cite{Sjostrand:2006za} for showering and hadronization. Finally simulation of the the detector level effects is performed using {\sc Delphes}~3 \cite{deFavereau:2013fsa}. The subsequent reconstruction of the events followed by a detailed cut-based analysis is performed using the {\sc MadAnalysis}~5 framework~ \cite{Conte:2012fm,Conte:2014zja}.

The dominant SM background which makes the experimental search challenging comes from $t\bar{t}$, $t\bar{t}$+jets, $Z$+jets and $ZZ$+jets process. We demand exactly two $b$ quarks are 2 leptons (either $e$ or $\mu$) in the final state. Understandably, this reduces our background considerably while also reducing our signal by an amount commensurate with the $b$-tagging efficiency at the LHC. We begin our kinematic analysis by imposing the following basic identification cuts on the final state particles:
\begin{equation}
p_T^j > 20~{\rm GeV}, \qquad p_T^\ell > 10~{\rm GeV}, \qquad
  |\eta^j|\leq5\qquad\text{and}\qquad  |\eta^\ell| \leq 2.5 \ .
  \label{eq:basic1}
\end{equation}

The basic identification cut on $p_{T}$ will help to eliminate the soft jets and leptons which arise during hadronization. We have deliberately chosen a wider window for pseudorapidity for jets as opposed to leptons so as to not lose too many signal events in the process. Further, to ensure that all pairs of final state particles are optimally separated, we impose the following separation cuts:   

\begin{equation}
\Delta R_{bb}=\Delta R_{ll}=\Delta R_{bl} \geq 0.4.
\label{eq:basic2}
\end{equation}

\begin{figure}[h!]
\includegraphics[scale=0.4]{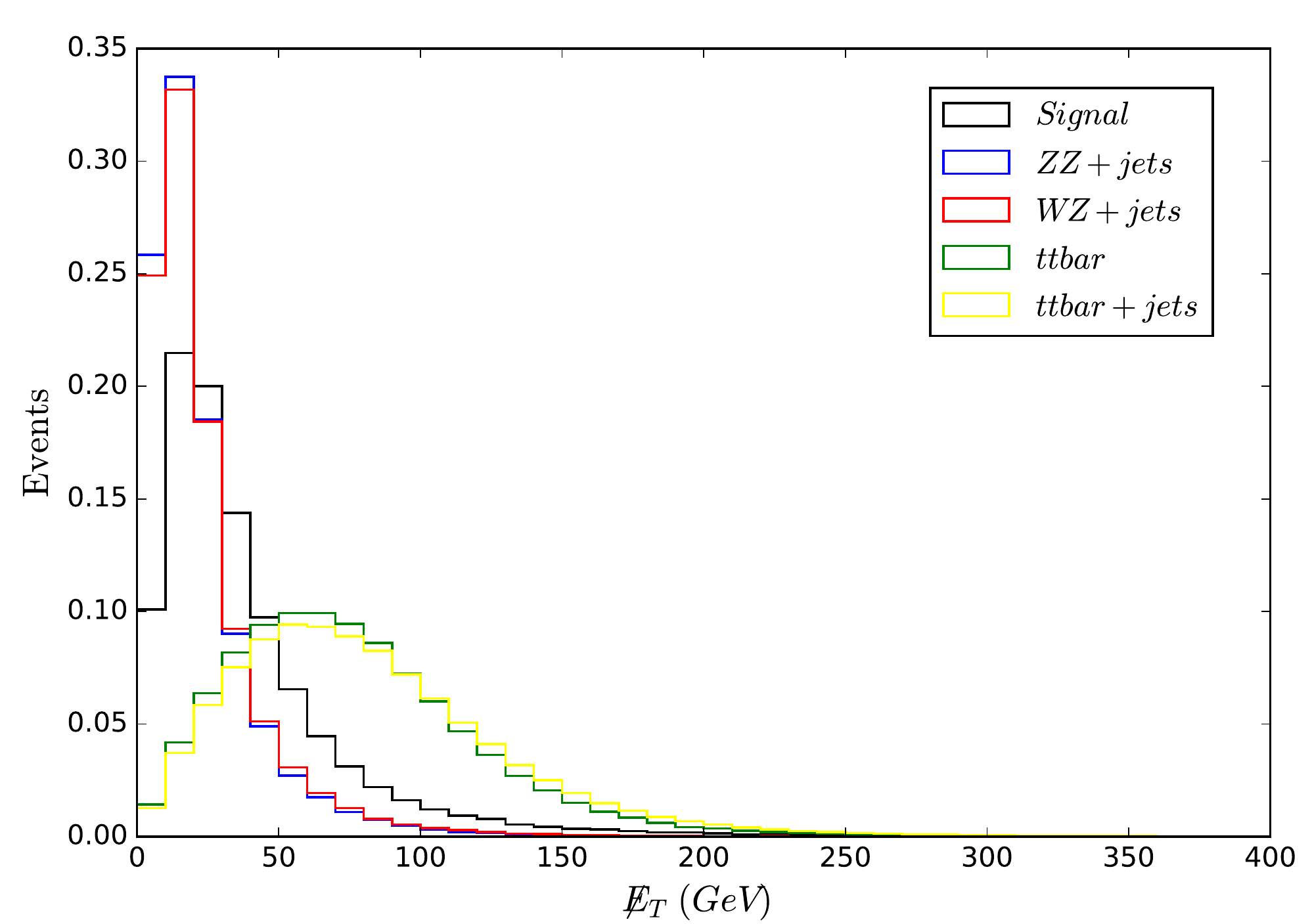}
\caption{The $\met$ distribution for both the signal and the background. It is seen that $t\bar{t}$ has a lot events with high missing energy and this enables us to choose an optimal cut to suppress the SM background. The benchmark point chosen for generating the signal distribution is $M_{Z'}=$ 500 GeV.} 
\label{fig:met}
\end{figure}

At the outset, it is important in any search with $t\bar{t}$ as a final state to optimize the S/B so the enormous production cross-section of the top quark does not nullify any meaningful signal. We resort to the fact that the signal in our case - $pp\to Z'\to Zh\to \ell\ell b b$ does not carry any significant source of missing energy. We therefore first look at $\met$ distributions to decide on an optimal cut -- we display this in Fig.~\ref{fig:met}. We choose $\met<$30 GeV to eliminate a significant portion of the $t\bar{t}$ background - it can be seen from Table \ref{tab:Z500} that this cut eliminates more than 85\% of the $t\bar{t}$ and $t\bar{t}$+jets backgrounds that remain after the basic identification cuts.

\begin{figure}[h!]
\includegraphics[scale=0.4]{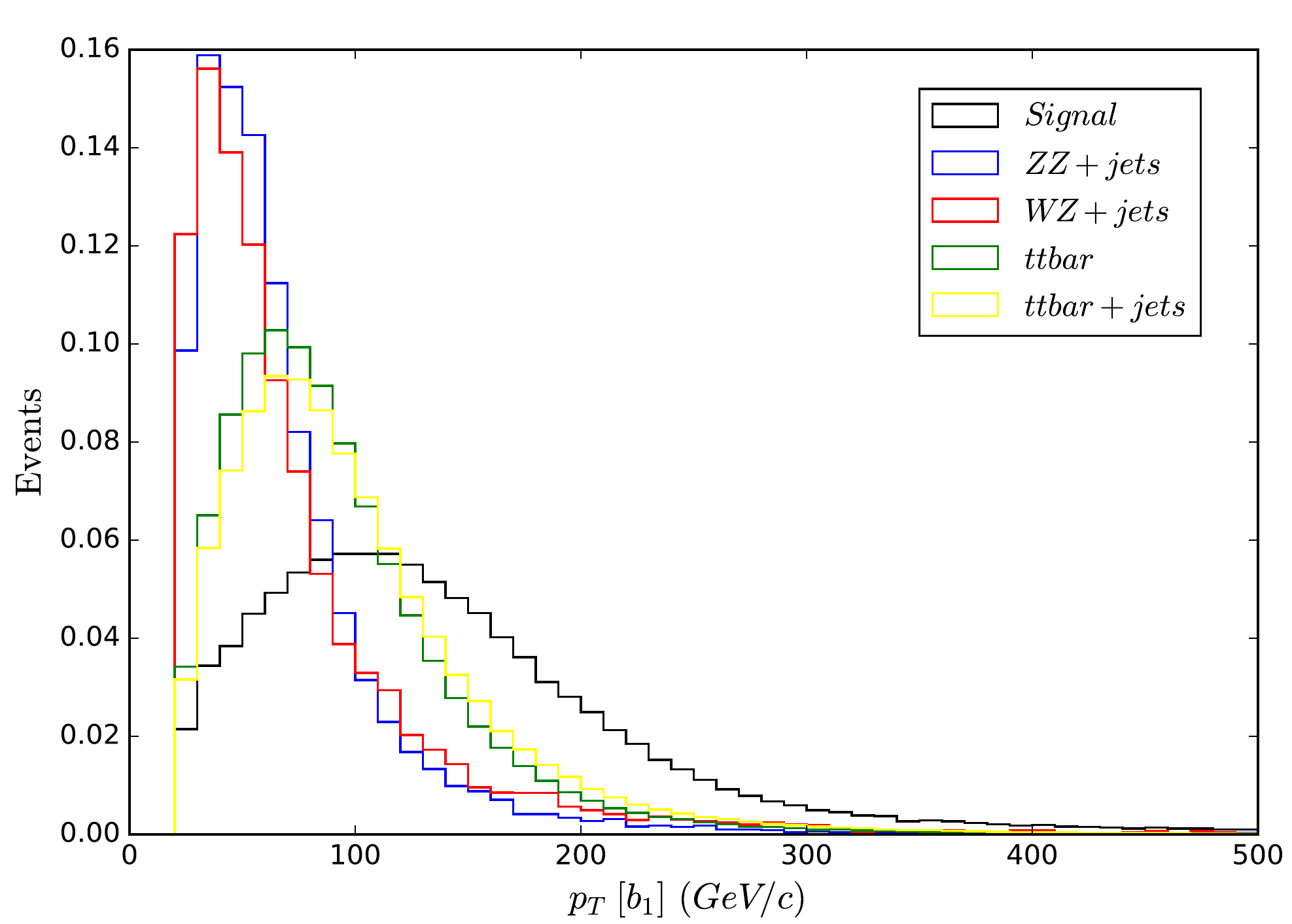}
\hspace{0.1in}
\includegraphics[scale=0.4]{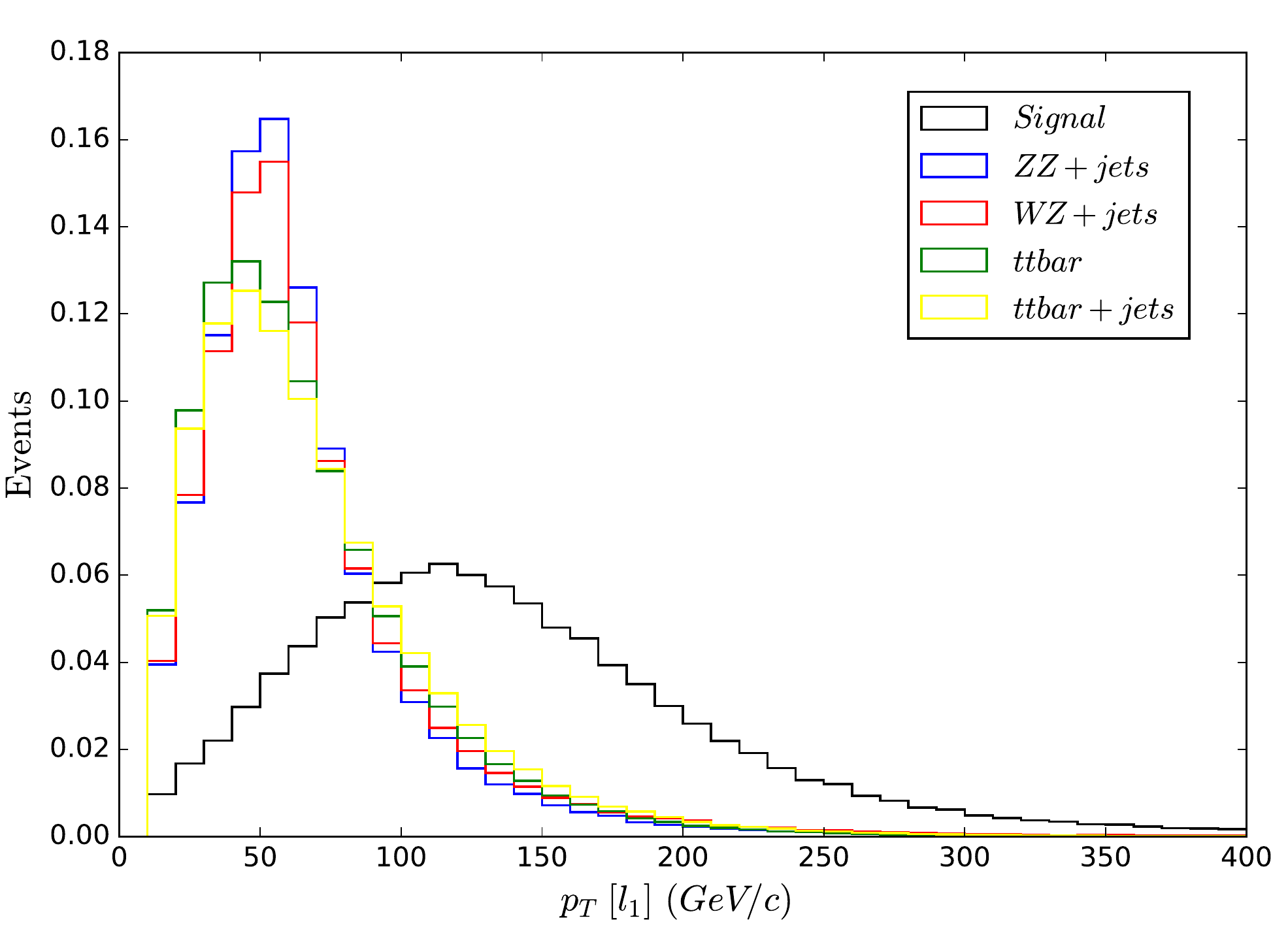}
\caption{The $p_{T}$ distribution of the leading $b$ jet and lepton for both the signal and background. The benchmark point chosen for generating the signal distribution is $M_{Z'}=$ 500 GeV.} 
\label{fig:pT}
\end{figure}

In Fig.~\ref{fig:pT}, we display the transverse momenta of the leading $b$ jet and lepton for both the signal (corresponding to $M_{Z'}=$ 500 GeV) and the SM background. We find that there is no need to impose strong cuts of the momenta of the final state particles -- the combination of $\met$ and the various invariant mass cuts conspire to reduce the SM significantly -- however we present these distributions so one has an idea of the magnitudes involved. Specifically, it is seen that the signal produces a significant fraction of hard $b$ jets with $p_T>$ 100 GeV as one would expect of the particles coming from the decay of a heavy $Z'$ of mass 500 GeV.

\begin{figure}[h!]
\includegraphics[scale=0.4]{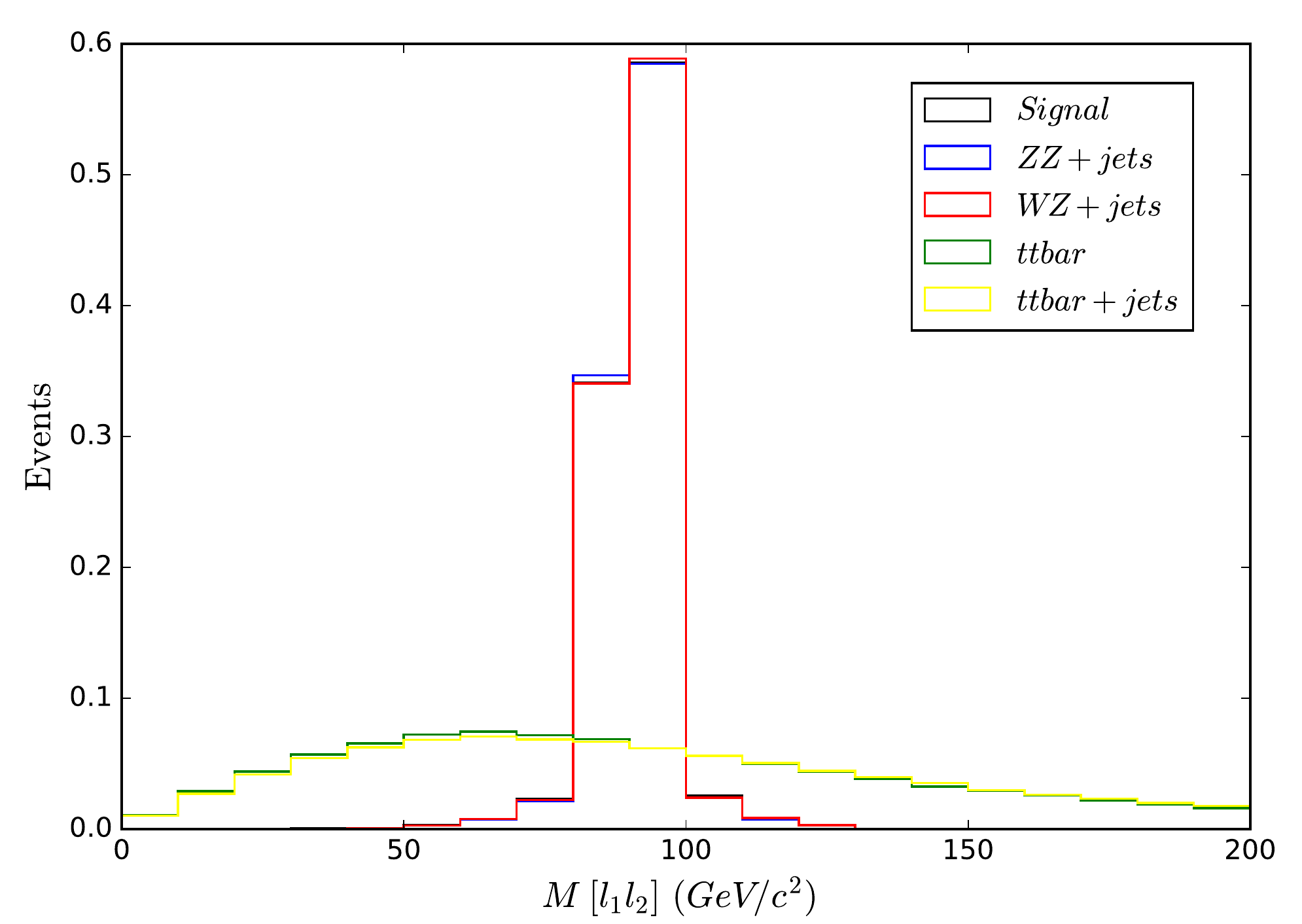}
\hspace{0.1in}
\includegraphics[scale=0.4]{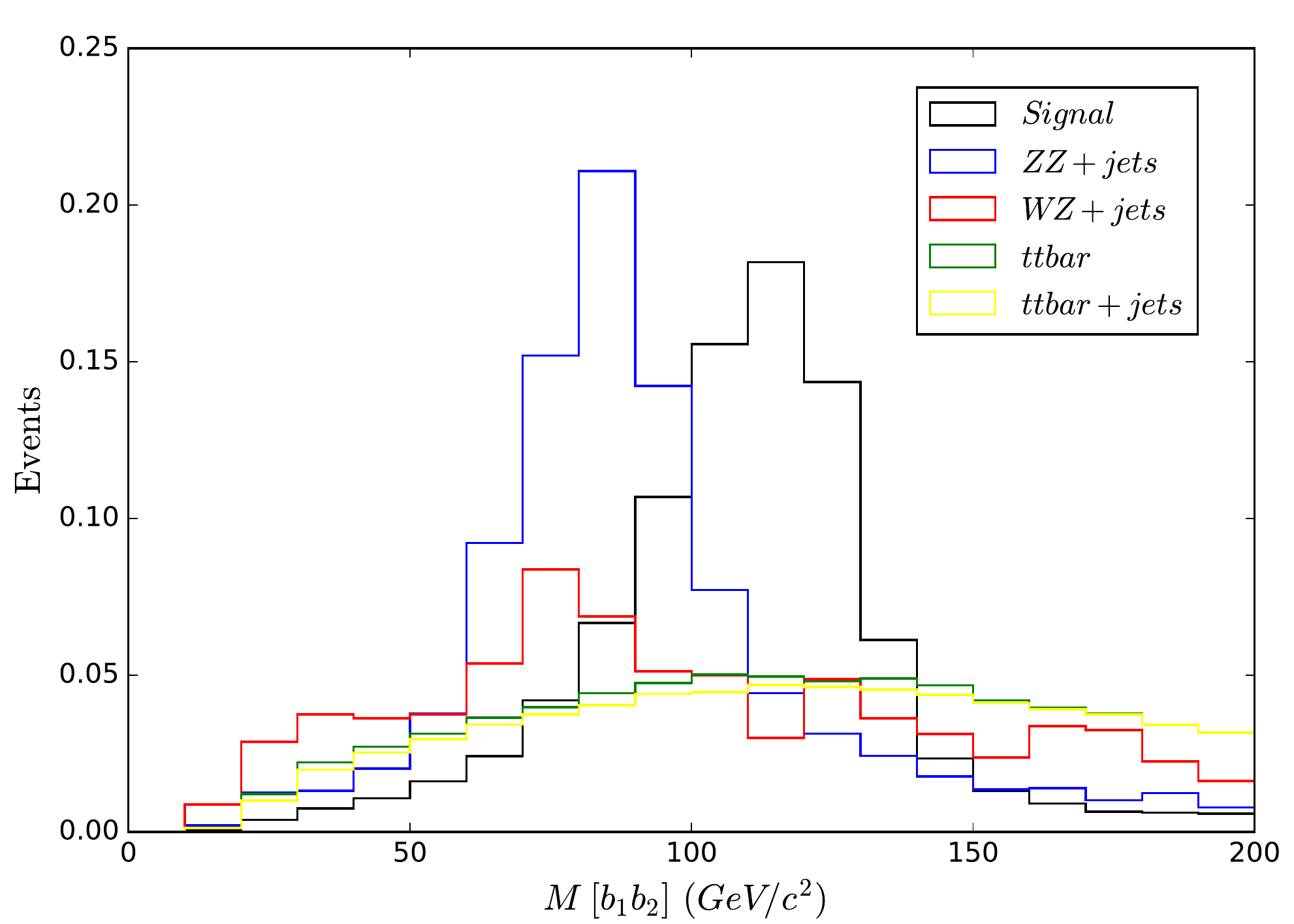}
\caption{The invariant mass distributions $M_{\ell_1\ell_2}$ and $M_{b_1b_2}$ for both the signal and background. It is seen that the remaining top quark background can be significantly eliminated by requiring the two leptons to come from the decay of a $Z$. The benchmark point chosen for generating the signal distribution is $M_{Z'}=$ 500 GeV.} 
\label{fig:inv}
\end{figure}

The invariant mass distribution for the leptonic pair and the pair of $b$ quarks is displayed in Fig.~\ref{fig:inv}. In the $M_{l_{1}l_{2}}$ case, the signal and background overlap to a significant degree for the $WZ$ and $ZZ$ backgrounds as the leptons originate from a $Z$ decay in both cases, but this observable can be used as an important discriminant for the $t\bar{t}$ background as the leptons there are kinematically quite dissimilar with the signal. Unlike most of the background, the invariant mass distribution for $b$ jets peaks at the mass of the SM Higgs - the distribution in this case is seen to be a little broad compared to the leptonic invariant mass. However, choosing a wider window for $M_{bb}$, we can still eliminate a substantial amount of SM background.
 
 Based on the lessons gleaned from the above plots, we choose the following set of invariant mass cuts to filter out the signal from the SM background:
\begin{equation}
80~\textrm{GeV} \leq M_{l_1l_2} \leq 100~\textrm{GeV};\,100~\textrm{GeV} \leq M_{b_1b_2} \leq 140~\textrm{GeV}.
\end{equation}
 
\begin{figure}[h!]
\includegraphics[scale=0.45]{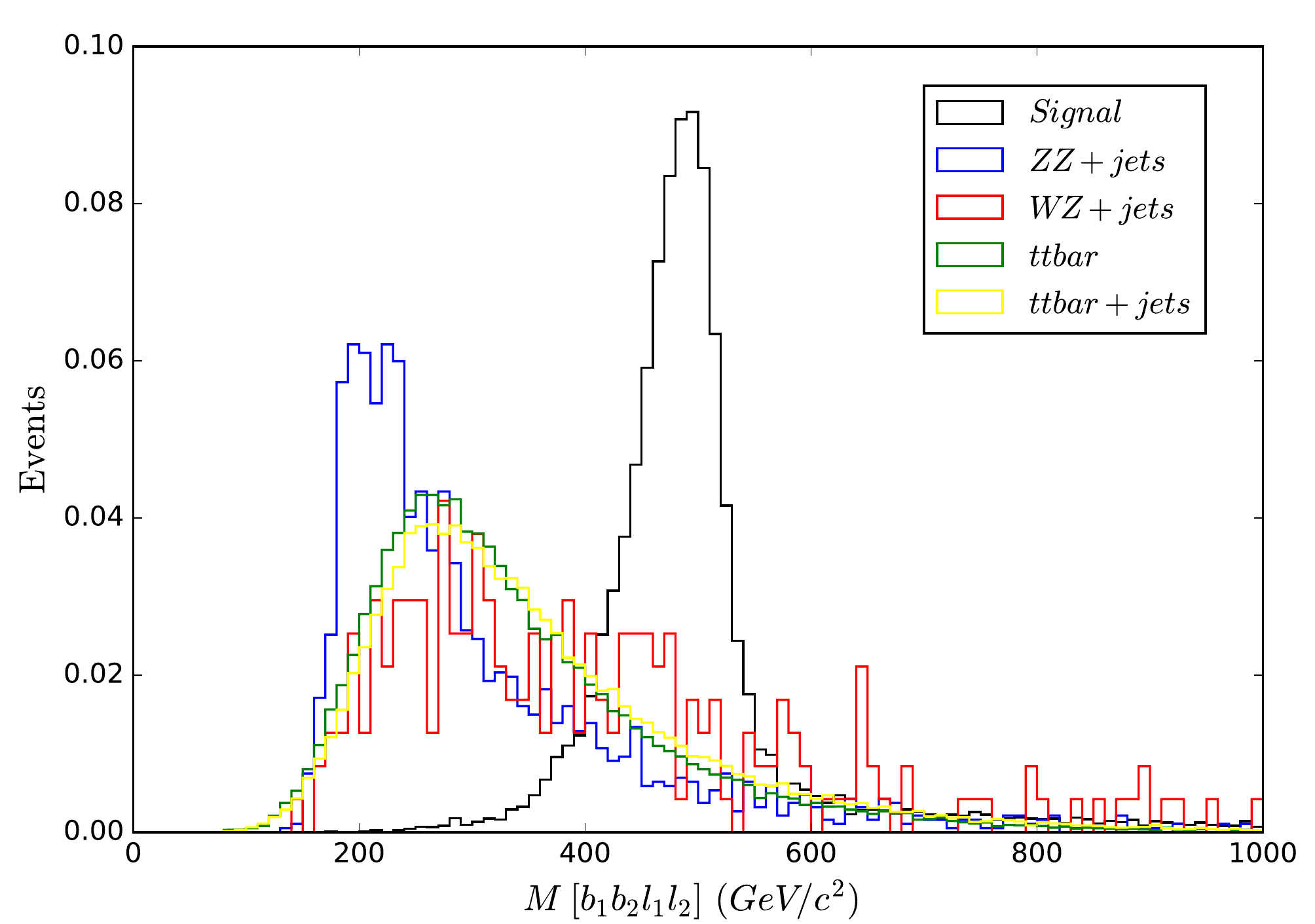}
\caption{The invariant mass distribution $M_{bb\ell\ell}$ for both the signal (corresponding to $M_{Z'}=$ 500 GeV) and the background. It is seen that there is a clear demarcation in this kinematic observable between the SM and the signal over the relevant region of interest around $M_{Z'}=$ 500 GeV.}  
\label{fig:invM}
\end{figure}

Finally, we turn to the final step in the process of isolating the events in the $M_{bb\ell\ell}$ distributions that correspond to the decay of the heavy $Z'$ - we first display the distribution with both the signal and the SM backgrounds in Fig.~\ref{fig:invM}. It is clear that our choice of cut $450~\textrm{GeV} \leq M_{b_1b_2l_1l_2} \leq 550~\textrm{GeV}$ does a very good job of isolating the signal from the background.

Finally, we present in Table \ref{tab:Z500} the complete cut flow chart that details the impact of each of the kinematic cuts employed on both the signal and the various backgrounds.
\begin{table}[h!]
\centering
\begin{tabular}{|c|c|c|c|c|c|c|c|}
\hline
 Cut selection   &Signal  &$ZZ + Jets$  &$WZ + Jets$ &$t \bar t$ &$t \bar t j$  &S/B &S/$\sqrt{B}$ \\
\hline\hline
 Initial                                         &100000  &200000  &200000 &300000 &400000 & - & -  \\
$N_l$ = 2                                        &53017   &68975   &71198   &110348  &138337  &0.136 &85.019 \\
$N_b$ = 2                                        &16337   &3263    &423    &32842      &43160  &0.205 &57.87 \\
$\met \le 30~GeV$                            &8942   &2725    &285    &4424   &5309  &0.7017 &79.213 \\
80~GeV $\leq M_{l_1l_2} \leq$ 100~GeV           &8299    &2545     &266     &582      &688  &2.033 &129.91 \\
100~GeV $\leq M_{b_1b_2} \leq$ 140~GeV           &5345    &362     &51     &137      &144  &7.7 &202.89 \\
450~GeV $\leq M_{b_1b_2l_1l_2} \leq$ 550~GeV     &4902    &15       &7     &0      &2  &204.25 &1000.67 \\
\hline
\end{tabular}
\caption{Showing the cross-section estimation of signal and background at the center-of-mass energy of 14 TeV at the LHC, for the $M_{Z'}$ of 500 GeV.}
\label{tab:Z500}
\end{table}

\begin{figure}[ht!]
\includegraphics[scale=1.0]{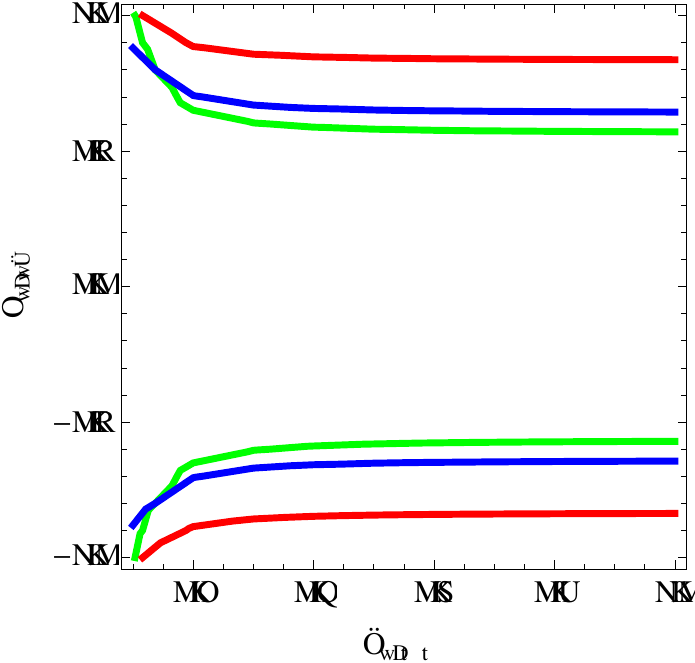}
\caption{The reach in the parameter space in the $g_{Z'Zh}-g_{Z'WW}$ plane for an integrated luminosity 500$fb^{-1}$ at the LHC. The red, blue, and green contours (colors online) show the parameter space in the $g_{Z'Zh}-g_{Z'WW}$ plane in which a 5$\sigma$ discovery of the $Z'$ is possible in the $\ell\ell b b$ final state for $M_{Z'}=$700, 500, and 300 GeV respectively. }
\label{fig:contour}
\end{figure}

Up until now in our analysis, we have chosen a fiducial signal cross-section to emphasize the efficacy of the cuts - the number chosen has no meaning as such as it is not a model-dependent analysis. Now, we turn to the question of how much signal cross-section one would need in a realistic model that would enable one to beat the SM backgrounds that remain after the imposition of all the cuts. We display in Table \ref{tab:csLHC}, the cross-section $\times$ BR needed in any specific BSM scenario with a fermiophobic $Z'$ to facilitate discovery in the $\ell\ell b b$ channel for various luminosities. While the numbers do not seem unreasonable, one still has to check whether they are indeed realizable in a specific model-dependent scenario. We perform that analysis in the context of the 221 model for an integrated luminosity of 500 fb$^{-1}$ at the LHC - our result is displayed in Fig.~\ref{fig:contour}. The red, blue, and green contours (colors online) show the parameter space in the $g_{Z'Zh}-g_{Z'WW}$ plane in which a 5$\sigma$ discovery of the $Z'$ is possible in the $\ell\ell b b$ final state for $M_{Z'}=$700, 500, and 300 GeV respectively. While larger values of $M_{Z'}$ would understandably be more difficult to probe owing to a reduction in the signal cross-section, there is a significant spread in the possible coupling values that aid discovery -- specifically almost the entire range of coupling values between 0 and 1.

\begin{table}[ht!]
\centering
\begin{tabular}{|c|c|c|c|}
\hline
$M_{Z'}$  & $\cal{L}$      & \multicolumn{2}{c|}{}         \\
              &                    &Background             &Signal                      \\
(GeV)   &($fb^{-1}$)      & (fb)                      &(fb)                               \\
\hline\hline
     &100  & &2.4   \\
300  &500     &28.69     &1.084 \\
     &1000    &     &0.767 \\
\hline
     &100  & &0.94   \\
500  &500     &1.8826     &0.42 \\
     &1000    &     &0.297 \\
   \hline
     &100  & &0.633   \\
700  &500     &0.1107     &0.283 \\
     &1000    &    &0.2 \\
\hline          
\end{tabular}
\caption{The signal cross section required for a 5$\sigma$ discovery of the $Z'$ in the $\ell\ell b b$ channel for various luminosities at the LHC.}
\label{tab:csLHC}
\end{table}

\subsection{$Z'$ search prospects at CLIC}
\label{subsec:CLIC}

In the last section we demonstrated that the future LHC which runs at a higher energy and increased luminosity provides a potential environment for the discovery of a fermiophobic $Z'$ in the $\ell\ell b b$ channel. Nevertheless, it is hard to ignore the various limitations a hadronic detector bears:  it is difficult to separate out the hadronic decays from the large QCD background in general, the flavor and energy of the initial state quark at the production level is hard to control and detect. Along with that, for our specific goal, we note that there is a steep fall in the parton distribution function limiting the accessibility of elementary VBF processes. By contrast a linear collider provides a much more clean environment where the VBF processes is not suppressed. The initial state is known in this case on an event by event basis and because electromagnetic radiation loss is minor for this collider, one has access to higher center of mass (C.M) energy scales. Compared to the hadronic collisions there are much more detailed complementary information is available for the leptonic collisions. With all these prospects,  the Compact Linear Collider (CLIC)\cite{Abramowicz:2013tzc} is an attractive option as a possible future multi-TeV linear electron-positron collider based on a novel dual-beam acceleration scheme designed to reach multi-TeV C.M energies. The CLIC potential allows to explore a very rich physics program with operations designed at three successive stages of higher C.M energies 380 GeV, 1.4 TeV and 3 TeV by achieving very high luminosities of 500 $fb^{-1}$, 1500 $fb^{-1}$ and 2000 $fb^{-1}$. We now turn to the prospect of discovering a fermiophobic $Z'$ in this collider -- we will only present a parton level analysis in this section.

\begin{figure}[!ht]
\begin{center}
\includegraphics[angle=0,width=70mm]{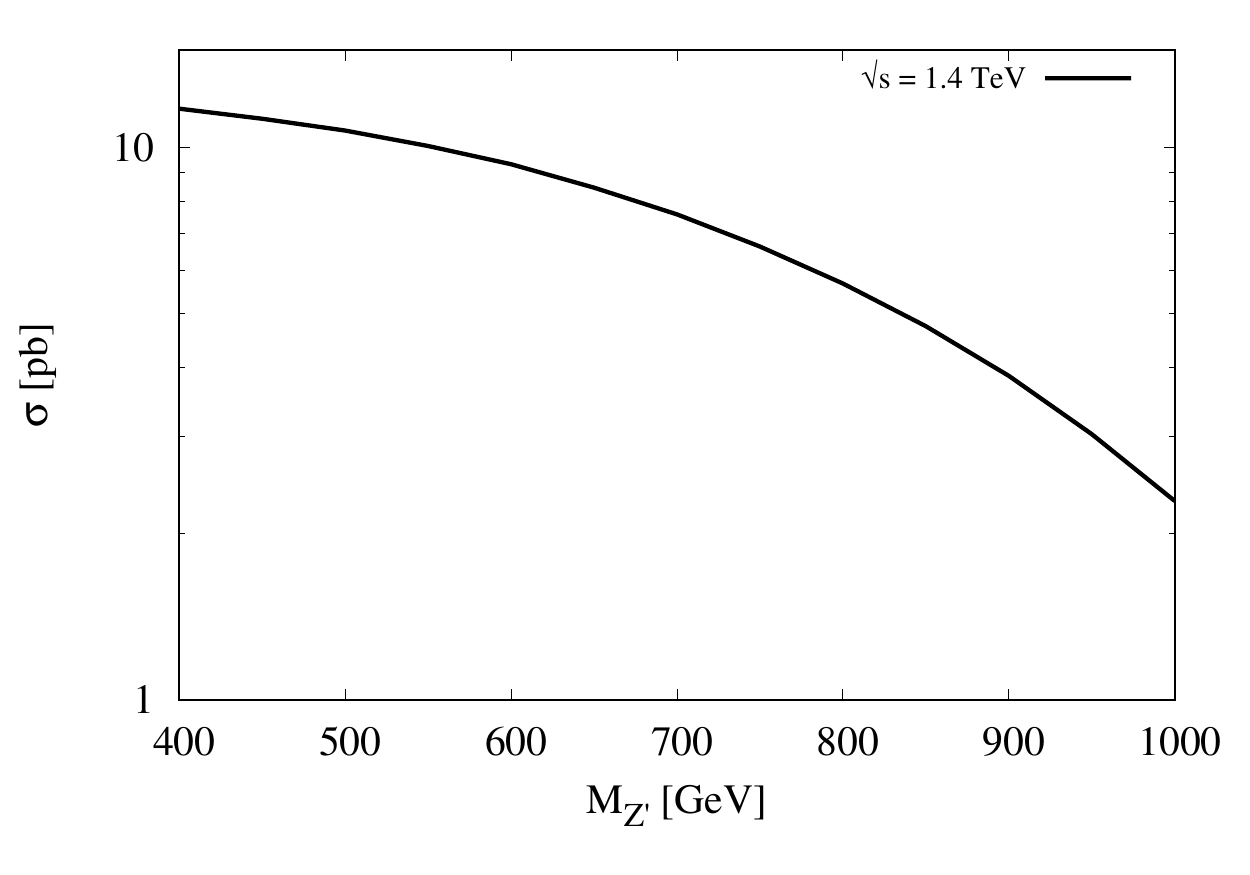}
\caption{The production cross-section as a function of $M_{Z'}$ for the $Z'$ production at the CLIC for a $\sqrt{s}=$1.4 TeV.}
\label{fig:CSCLIC}
\end{center}
\end{figure}

For our study we fix the C.M energy to be 1.4 TeV and choose an integrated luminosity at the corresponding energy stage 1500 $fb^{-1}$ at the CLIC. The heavy neutral boson is produced via $WW$ fusion and $Z'$ further decays into the SM neutral gauge boson $Z$ and the SM Higgs boson $h$. The complete process under consideration is thus $e^{-}e^{+} \rightarrow Z'\nu_{e}\bar{\nu}_{e} \rightarrow h Z \nu_{e}\bar{\nu}_{e} \rightarrow l^{+} l^{-} b \bar{b} \nu_{e} \bar{\nu}_{e}$. We perform our analysis at the parton level, where the signal and background event generation has been done by using the Monte-Carlo event generator package MadGraph5\_aMC. In Fig.~\ref{fig:CSCLIC}, we show the variation of the total cross-section as a function of the mass of the heavy neutral gauge boson $M_{Z'}$ -  as expected the value of $\sigma$ decreases with increasing $M_{Z'}$.

\begin{figure}[h!]
\begin{center}
\includegraphics[angle=0,width=80mm]{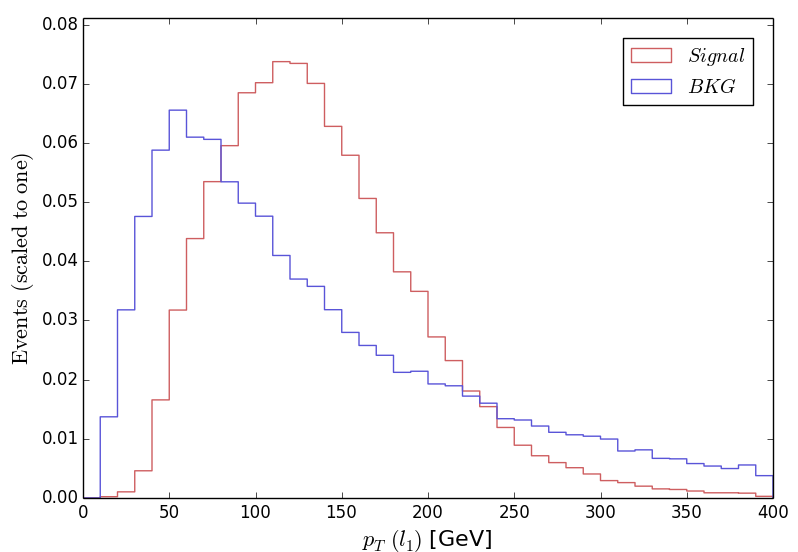}
\includegraphics[angle=0,width=80mm]{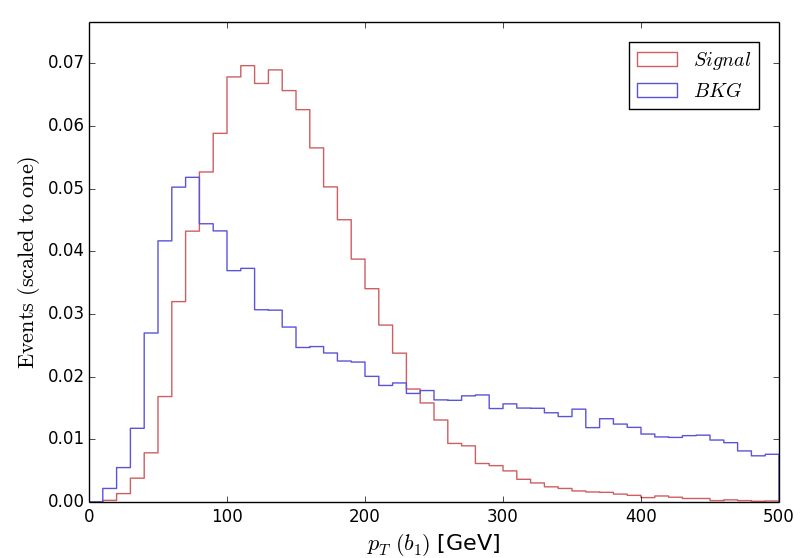}\\
\caption{The transverse momentum distributions of leptons and b-quarks for the $Z'$ mass of 300, 500 and 700 GeV (from left to right) respectively, at the fixed center-of-mass energy of 1.4 TeV.}
\label{fig:clic1}
\end{center}
\end{figure}

For our analysis we choose the same three signal benchmark points: $M_{Z'}=$ 300 GeV, 500 GeV and 700 GeV and generate the complete irreducible SM background for the process $l^{+} l^{-} b \bar{b} \nu_{e} \bar{\nu}_{e}$. We start by imposing an event selection cut of ($ N_{b} = 2, N_{l} = 2 $). This will reduce the SM background around 50\% without losing any signal events. The $p_T$ distributions of the leading  $b$ jet and lepton is shown in Fig.~\ref{fig:clic1} -- it is clear that there is a large portion of overlap between the signal and the SM background -- we thus put in a moderately high $p_T$ cut in order not to lose significant signal.    

We thus choose the following set of cuts:
\begin{equation}
p_T[b_1] \geq 40~{\rm GeV},\, p_T[b_2] \geq 20~{\rm GeV},\, p_T[l_1] \geq 40~{\rm GeV}, \,p_T[l_2] \geq 20~{\rm GeV}, \,\met \geq 40~{\rm GeV} .
  \label{eq:basic4}
\end{equation} 

\begin{figure}[h!]
\begin{center}
\includegraphics[angle=0,width=80mm]{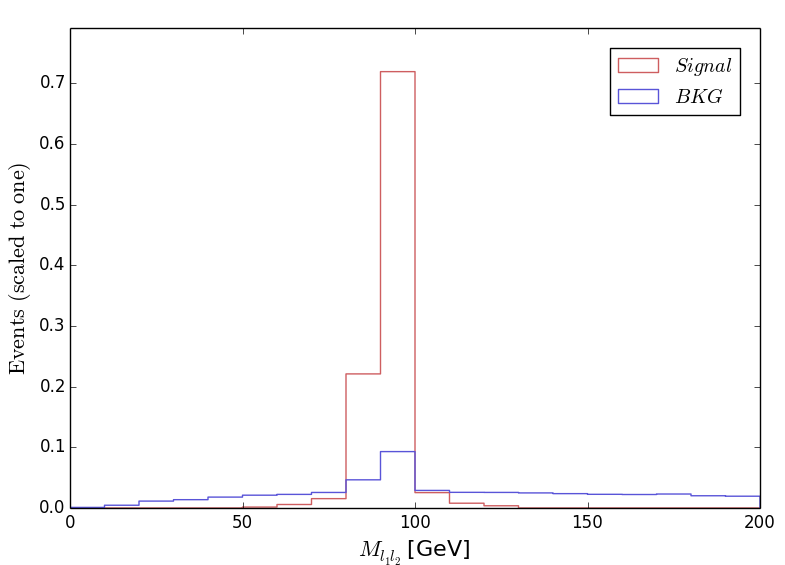}
\includegraphics[angle=0,width=80mm]{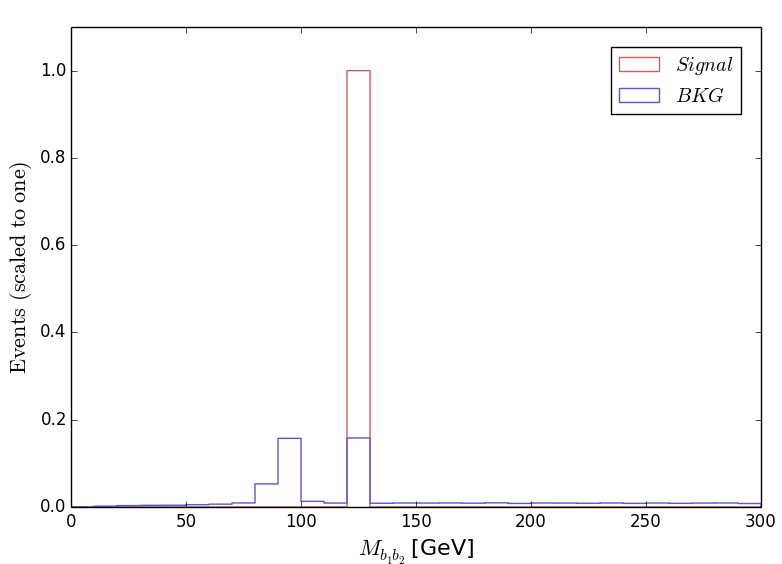}\\
\includegraphics[angle=0,width=80mm]{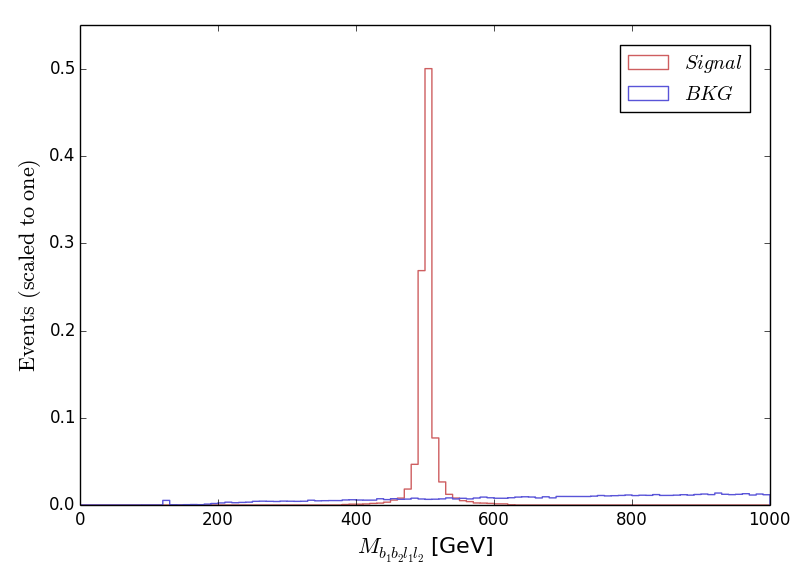}
\caption{The invariant mass distributions of $l_1l_2$, $b_1b_2$ and $b_1b_2l_1l_2$ for the $Z'$ mass of 300, 500 and 700 GeV (from left to right) respectively, at the fixed center-of-mass energy of 1.4 TeV.}
\label{fig:clic4}
\end{center}
\end{figure}

The invariant mass distributions $M_{\ell\ell}$, $M_{bb}$, and $M_{\ell\ell b b}$ are presented in Fig.~\ref{fig:clic4}.  Of course, this being a parton level analysis, we obtain relatively sharp peaks for all the distributions. Detector effects will undoubtedly smear these plots by some amounts. The sharp well-defined peak in the signal distribution enables us to choose a tight invariant mass window:
\begin{equation}
(m_{Z^{'}} - 50) GeV \leq m_{l_{1}l_{2}b_{1}b_{2}} \leq (m_{Z^{'}} + 50) GeV
\end{equation}
to eliminate the SM background and improve the $S/\sqrt{B}$ efficiency.

Finally, in Table \ref{tab:csCLIC}, we present the cut flow chart detailing the effects of the various kinematic cuts -- as before, we have chosen a fiducial cross-section for the signal and one needs to understand whether the remaining background can be suitably masked by the signal in a specific model. However, the number of background events that remain after all the cuts are imposed for the $M_{Z'}=$ 500 GeV case is around 80, which would mean that one needs around 60 signal events for a 5$\sigma$ discovery. For the operating luminosity of 1500 fb$^{-1}$, this translates to a signal cross-section of 0.04 fb. Even for highly suppressed branching ratios, it is clear from Fig.~\ref{fig:CSCLIC} that the production cross-section is large enough to aid the discovery. A full detector level analysis would undoubtedly push the required signal cross-section to higher numbers, however the present analysis serves as proof of concept that the CLIC offers an extremely good opportunity to unravel new physics in extended gauge models.

\begin{table}[h!]
\centering
\begin{tabular}{|c|c|c|c||c|c|c|}
\hline
 Cut selection    & \multicolumn{3}{c|}{Signal at $M_{Z'}$}        & \multicolumn{3}{c|}{Background}       \\
                &300 GeV  & 500 GeV       &700 GeV                 &                     &  &              \\
                &(S1)      &(S2)          &(S3)                    &(S1)                 &(S2)  & (S3)     \\
\hline\hline
 Initial                                          & 50000 $\pm$ 0     & 50000 $\pm$ 0     & 50000 $\pm$ 0      & 50000 $\pm$ 0  &50000 $\pm$ 0 &50000 $\pm$ 0   \\
\hline
$N_b$ = 2.0                                       & 50000 $\pm$ 0     & 50000 $\pm$ 0     & 50000 $\pm$ 0      & 28234 $\pm$ 110 &28234 $\pm$ 110 &33847 $\pm$ 104   \\
$N_l$ = 2.0                                       & 50000 $\pm$ 0     & 50000 $\pm$ 0     & 50000 $\pm$ 0      & 28234 $\pm$ 110 &28234 $\pm$ 110 &28234 $\pm$ 110   \\
$MET \ge 40~GeV$                                  & 46291 $\pm$ 59    & 44502 $\pm$ 70    & 42418 $\pm$ 80     & 24696 $\pm$ 111 &24696 $\pm$ 111 &24696 $\pm$ 111   \\
$P_T(b_1) \ge 40~GeV$                             & 44878 $\pm$ 68    & 44247 $\pm$ 71    & 42296 $\pm$ 81     & 22724 $\pm$ 111 &22724 $\pm$ 111 &22724 $\pm$ 111   \\
$P_T(b_2) \ge 20~GeV$                             & 36654 $\pm$ 99    & 38200 $\pm$ 95    & 38397 $\pm$ 94     & 19620 $\pm$ 109 &19620 $\pm$ 109 &19620 $\pm$ 109   \\
$P_T(l_1) \ge 40~GeV$                             & 36005 $\pm$ 100   & 37990 $\pm$ 96    & 38272 $\pm$ 95     & 19320 $\pm$ 108 &19320 $\pm$ 108 &19320 $\pm$ 108   \\
$P_T(l_2) \ge 20~GeV$                             & 32202 $\pm$ 107   & 33498 $\pm$ 105   & 33847 $\pm$ 104    & 17833 $\pm$ 107 &17833 $\pm$ 107 &17833 $\pm$ 107   \\
85~GeV $\leq M_{l^-l^+} \leq$ 100~GeV            & 29501 $\pm$ 109   & 30561 $\pm$ 109   & 30928 $\pm$ 108    & 1292 $\pm$ 36   &1292 $\pm$ 36   &1292 $\pm$ 36     \\
100~GeV $\leq M_{b_1b_2} \leq$ 140~GeV           & 29501 $\pm$ 109   & 30561  $\pm$ 109  & 30928 $\pm$ 108    & 391  $\pm$ 20   &391  $\pm$ 20   &391  $\pm$ 20     \\
$(M_{Z'}-50) \leq M_{b_1b_2l^-l^+} \leq (M_{Z'}+50)$ & 29501 $\pm$ 109  & 29687  $\pm$ 109  & 27532 $\pm$ 111    & 108   $\pm$ 10    &78   $\pm$ 9    &15   $\pm$ 4   \\
\hline\hline
\end{tabular}
\caption{Cut flow chart displaying the efficiencies of each kinematic cut chosen -- the analysis is done for a center-of-mass energy of 1.4 TeV in the context of CLIC.}
\label{tab:csCLIC}
\end{table}

\section{Conclusions}
\label{sec:conclusions}
Most extensions of the SM fall into a few broad classes: ones enlarging the gauge sector of the theory, ones enlarging the scalar sector (and thus the EWSB structure), and ones that enlarge the matter content. Of course, many models fall under more than one category. Given this proliferation in the model building scene, it is important for theorists and phenomenologists alike to look for common cues in many of these scenarios -- this serves to both discover new physics at the LHC (and other colliders) and also to look for distinguishing features that would help the inverse program of mapping from a potential future discovery to the space of models. In this paper, we considered a class of models that are characterized by an enlarged gauge spectrum with an additional heavy $Z'$ present in the low energy theory which is fermiophobic in nature. While our analysis has been model independent, we made use of the 221 model in the literature to help us translate our results into the parameter space of the theory.

We analyzed the discovery prospects of the fermiophobic $Z'$ in the process $pp\to Z'\to Zh\to\ell\ell b b$ at the 14 TeV LHC. We generated both the signal and the complete SM backgrounds, and systematically put in various kinematic cuts to reduce the SM background. We find that to discover a heavy $Z'$ of mass 500 GeV (700 GeV) in the $\ell\ell b b$ final state at the LHC, one needs a signal cross-section of 0.42 fb (0.28 fb) at an integrated luminosity of 500 fb$^{-1}$. We translated these numbers into the parameter space of a specific 221 model and found that a large combination of the coupling values $g_{Z'WW}$ and $g_{Z'Zh}$ in the range between 0 and 1 allow for a 5$\sigma$ discovery of the $Z'$ with the process under study. While this is encouraging, we also undertook a preliminary study of the discovery process of the $Z'$ at the future CLIC linear collider looking at the process $e^{-}e^{+} \rightarrow Z'\nu_{e}\bar{\nu}_{e} \rightarrow h Z \nu_{e}\bar{\nu}_{e}$ with the $Z$ decaying leptonically and the higgs decaying to $b\bar{b}$. We performed a parton level analysis and find that the values of $\sigma\times$BR required for a 5$\sigma$ discovery can be amply provided in many models even with suppressed $Z'\to Z h$ branching ratios. 

While the search for $Z'$ and all BSM scenarios is going on in full force at the LHC, it is right time to look for signatures that might be hidden from us in cases where the conventional search channels do not apply for a specific class of models. This paper summarizes the search strategy that one could employ to discover fermiophobic $Z'$s that could be part of the spectrum of a class of models.

\begin{acknowledgments}
BC would like to acknowledge the support by the Department of Science and Technology under Grant YSS/2015/001771.  SK acknowledges financial support from the Department of Science and Technology, India, under the National Post-doctoral Fellowship programme, Grant No. PDF/2015/000167. 
\end{acknowledgments}

\end{document}